\documentclass[twocolumn,english,prl,superscriptaddress,showpacs,nobalancelastpa
ge]{revtex4}
\usepackage[latin1]{inputenc}

\makeatletter

\usepackage{graphicx}
\usepackage{bm}
\usepackage{babel}

\makeatother
\begin{document}

\title{Coherence-preserving trap architecture for long-term control of giant
Rydberg atoms}

\author{P. Hyafil}

\author{J. Mozley}

\author{A. Perrin}

\author{J. Tailleur}

\author{G. Nogues}

\author{M. Brune}

\author{J.M. Raimond}

\affiliation{Laboratoire Kastler Brossel, D\'{e}partement de Physique de l'Ecole
Normale Sup\'{e}rieure, 24 rue Lhomond, F-75231 Paris Cedex 05, France}

\author{S. Haroche}

\affiliation{Laboratoire Kastler Brossel, D\'{e}partement de Physique de l'Ecole
Normale Sup\'{e}rieure, 24 rue Lhomond, F-75231 Paris Cedex 05, France}

\affiliation{Coll\`{e}ge de France, 11 place Marcelin Berthelot, F-75231 Paris
Cedex 05, France}

\date{\today{}}

\begin{abstract}
We present a way to trap a single Rydberg atom, make it long-lived
and preserve an internal coherence over time scales reaching into
the minute range. We propose to trap using carefully designed
electric fields, to inhibit the spontaneous emission in a non
resonant conducting structure and to maintain the internal
coherence through a tailoring of the atomic energies using an
external microwave field. We thoroughly identify and account for
many causes of imperfection in order to verify at each step the
realism of our proposal.
\end{abstract}


\maketitle

Giant circular Rydberg atoms are blessed with remarkable
properties \cite{TXT_GALLAGHER}. They have a long lifetime and a
huge electric dipole. Their consequent strong coupling to static
or resonant electric fields makes them highly sensitive probes of
their environment. In recent years, they have been used, coupled
to millimetre-wave superconducting cavities, to test fundamental
quantum concepts \cite{ENS_FUNDAMENTAL} and to realize elementary
quantum logic operations \cite{ENS_RMP,ENS_COLLISION}. 
In this context, Rydberg atoms are competing with other systems, such as ions
\cite{ION_BLATTWINELANDREVIEW03} or neutral atoms \cite{QC_BLOCH03}, used or
proposed, under various conditions, as tools for 
the testing of quantum information procedures. A new trend in this field
proposes the integration of atomic and mesoscopic systems allowing for new
applications of quantum mechanics
\cite{QI_LUKINTRANSMISSIONLINE04,TR_SCHOELKOPFTRANSMISSIONLINE04}. In this
regard, Rydberg atoms are very
promising candidates, provided their large sensitivity to electric
perturbations can be combated. They have the particular advantage of interacting
strongly with microwave fields which can be generated by high frequency
electronic circuits.  

In this Letter, we describe a general procedure making it possible to trap
and manipulate single Rydberg atoms. A circular Rydberg atom-chip trap design
stores individual atoms, prepared on demand. This cryogenic electric trap,
preventing spontaneous emission, makes circular states stable for minutes.
Moreover, it allows for the preparation, preservation and probing
of a millimetre-wave coherence between two levels. More complex
architectures of traps, guides, cavities and detectors could be
implemented on a single chip. This paves the way
to very high resolution atom/surface and atom/atom interaction
studies, to the coherent coupling of giant Rydberg atoms with
superconducting mesoscopic circuits, to complex quantum
entanglement knitting and to scalable quantum information
operations.

We discuss, for the sake of definiteness, the trap in the specific case of two
circular levels $e$ and $g$ with principal quantum numbers $n=51$ and $50$
respectively. These levels have maximum orbital ($l$) and magnetic ($m$) quantum
numbers ($l=|m|=n-1$).
The $e/g$ transition is at 51.099 GHz. Both circular levels have a long
radiative lifetime $T_{sp}=\Gamma_{sp}^{-1}\equiv30$~ms. Their preservation
requires a static electric (or magnetic) field
. This defines a physical quantization axis (unit vector $\bm{u}$) and
prevents mixing with other states in the hydrogenic manifold.

The radiative lifetime, although long, can be greatly increased by
inhibiting spontaneous emission (S.E.) \cite{ENS_HOUCHES90}. Both
levels $e$ and $g$ have a single decay channel, the emission of a
millimetre-wave photon circularly polarized in a plane
perpendicular to $\bm{u}$ (`$\sigma^\pm$ polarization'). This
emission is blocked when the atom is placed between two parallel
conducting planes (normal to $Oz$, itself ideally parallel to
$\bm{u}$) separated by a distance $d < \lambda/2$ ($\lambda \sim
6$ mm being the emitted photon wavelength). Lifetime increases by
a factor of a few tens have been demonstrated in the
millimetre-wave domain \cite{QC_INHIBITION}. Levels $e$ and $g$ can
simultaneously be made long-lived. There are two main
contributions to the residual spontaneous emission rate $\Gamma$.
The first, $\Gamma_a=\Gamma_{sp} \sin^2 \theta$, is due to any
eventual angle $\theta$ between $Oz$ and $\bm{u}$. The second,
$\Gamma_s$, is due to imperfections in the conducting surfaces.
Both these rates are enhanced by thermal processes and it is thus
essential to operate at cryogenic temperatures, below 1K. It is
important to note that, even in this S.E.-inhibiting structure,
the $e/g$ transition can be probed by means of an evanescent field
produced by an intense source. Note also that the atom-surface
distance (0.5 mm in the trap considered here) is
large enough that van der Waals interactions are but a small
perturbation.

Taking advantage of this extended lifetime requires a trapping
mechanism compatible with S.E. inhibition. We propose here an
electric trap relying on the quadratic Stark polarizability of $e$
and $g$ [$\alpha \equiv -0.2$ kHz/(V/m)${}^2$ for weak fields] in
a field $\bm{E}$ approximately aligned with $Oz$ (unit vector
$\bm{u_z}$). Since circular states are high-field seekers, there
can be no static field trap. We therefore turn to a dynamic
scheme, reminiscent of the Paul
trap \cite{ION_BLATTWINELANDREVIEW03}, combining static and
oscillating fields. Our design is comparable to that discussed in
Ref.~\cite{TR_PEIK99} for ground state atoms. 
Our trap geometry is presented in Fig.~1(a). The
S.E.-inhibiting planes are now made up of concentric electrodes. The
static potential $U_0$ creates an homogeneous directing
field $E_0\bm{u_z}$. The potential $U_1$, oscillating at
$\omega_1$, creates a smaller, a.c., approximately hexapolar field $\bm{E_1}$
[see Fig.~1(b)]. The atomic Stark energy, $-\alpha(E_0\bm{u_z}+ \bm{E_1})^2$,
has
a roughly quadratic spatial dependence. In order to cancel $\bm{E_1
}$ at the origin $O$ for all times, we apply the potential $\pm
\eta U_1$ to the outer electrodes, the factor
$\eta$ being determined by the electrode geometry.
Finally, the yet smaller static potential $U_2$ creates an
approximately quadrupolar field. This provides a force, nearly
constant in the trap region, compensating gravity (antiparallel to
$Oz$).

\begin{figure}
\includegraphics[width=8cm]{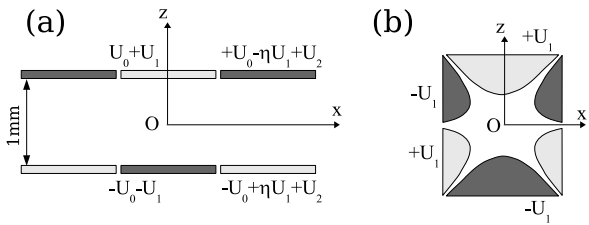}
\caption{\label{Fig:geometry} (a) Section of the trap in a vertical $xOz$
plane, with applied potentials. The trap has a cylindrical symmetry around $Oz$.
The diameter of the inner electrode is 1~mm. The plate spacing, 1~mm, is
appropriate for spontaneous emission inhibition. The electrodes are shaded
according to the phase of the oscillating potential $U_1$. (b) Electrode
geometry creating an exact hexapolar potential. The analogy with the
geometry of (a) is conspicuous. }


\end{figure}

The trap performance is assessed through numerical calculations. In order to
account for edge effects, we compute the
electric field using SIMION software. The corresponding atomic energies are
derived from the hydrogenic Stark formulae up to second order. The atomic
trajectories are computed using an adaptive-step Runge-Kutta
method. We set initial conditions corresponding to an atomic cloud
in a magnetic micro-trap with an adjustable temperature $T_0$.
For $U_0=$0.2~V, $U_2=$-0.003~V and an hexapolar potential $U_1=$0.155V
oscillating at $\omega_1=2\pi\times$430~Hz, the resulting motion is the
combination of a fast micromotion, at frequency $\omega_1/2\pi$, with a slow
oscillation whose longitudinal (along $Oz$) and transverse (orthogonal to $Oz$)
frequencies are 175 and 64~Hz respectively. For such low frequencies, the
orientation of the circular orbit adiabatically follows the space- and
time-dependent direction $\bm{u}$ of the local electric
field \cite{ENS_DAHU}. At $T_0$=90~$\mu$K, the motion has an extension
of around 100~$\mu$m. We have checked that for $T_0>100$~nK, the atomic
excursion in
the Rydberg trap is much
larger than the de Broglie wavelength, allowing a classical
treatment. The trap depth, $T_d$, which we define as the $T_0$ value for which
half of
the atoms remain within 400~$\mu$m of the origin, is $T_d =180$ $\mu$K, well
within the reach of standard laser cooling techniques.  We here use
potentials in the Volt range (much lower than in \cite{TR_PEIK99}), compatible
with the coherence-preservation scheme detailed below. Deeper traps are
achievable with higher voltages. 
The electrode planes efficiently inhibit
S.E. in the trap since the mean angle $\theta$ between $\bm{u}$
and $Oz$ is only 10~mrad for atoms at 90~$\mu$K, corresponding to
a residual lifetime of $1/\Gamma_a$=300~s.

Deterministic loading of the trap with a single Rydberg atom can be realized by
the `dipole blockade' effect \cite{QI_LUKINDIPOLEBLOCKADE01}.
This ensures the preparation of one and only one
low-angular momentum Rydberg state by laser excitation
 \cite{TR_SAFFMANN02}. The level shifts due to the Dipole-Dipole
interaction (1 GHz for two $n=50$ atoms separated by 1 $\mu$m) make the
laser non-resonant for a second Rydberg excitation. 
A correctly designed laser excitation chain could perform
a '$\pi$-pulse', whose
duration could be as short as a few $\mu$s,  preparing exactly one Rydberg atom
with a probability close to 1 \cite{TR_SAFFMANN02}.
This single atom can then be
transferred, within 20~$\mu$s, into the circular state by an adiabatic
process \cite{ENS_RMP}.
The short duration of this sequence allows for re-capture in the electric trap
before significant drifting has occurred. To create the initial high density
atomic cloud, we plan to work with ground-state atoms magnetically confined on
an atom-chip \cite{MX_HANSCHCHIP99}
which can provide submicron-sized samples at temperatures $T_0$ as low
as a few hundred nK \cite{MX_HANSCHBECCHIP01}. 
A double layer configuration would allow superposition of the magnetic and the
electrodynamic traps, with gold electrodes on top of the
current-carrying superconducting wires necessary for the magnetic trap. The gold
layer will efficiently inhibit spontaneous
emission, while not significantly perturbing the trapping
magnetic field created by the wires of the lower layer. This field is switched
off upon Rydberg state excitation. 

The Stark polarizabilities of levels $e$ and $g$ being slight\-ly dif\-fe\-rent,
the $e/g$ transition is dramatically broadened in the electric field trap. An
atom, prepared from a cloud at
0.3 $\mu$K, experiences a mean electric field amplitude, $E_a=400$ V/m,
with typical variations of $\pm 1$~V/m over its trajectory. Fig. 2(a)
shows the energy levels as a function of the electric field amplitude. Atomic
motion results in a 20~kHz broadening of the transition. An $e/g$ coherence
therefore decays within a few tens of $\mu$s. In addition,
the trajectories for $e$ and $g$ are rapidly separated (`Stern Gerlach' effect
due to the different trap potentials experienced by $e$ and $g$).  Coherence is
lost when this separation exceeds the wave-packet coherence length, of
the order of the de Broglie wavelength (here, about 0.8~$\mu$m).

\begin{figure}
\includegraphics[width=7cm]{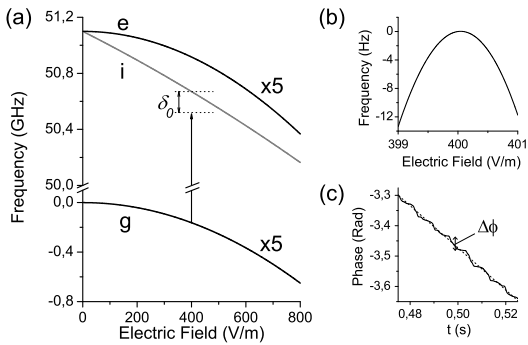}
\caption{\label{Fig:compensation} (a) Stark
energy of levels $e$, $g$ and $i$ versus electric field. States $g$ and $e$ have
a small, approximately quadratic Stark effect, magnified by a factor 5 for the
sake of
clarity. Level $i$ has a much larger, linear Stark effect. The microwave
dressing (vertical arrow) predominantly mixes $g$ and $i$ and reduces the
difference in Stark polarizability between $e$ and $g$. (b)
Relative frequency $\omega_{eg}(E)-\omega_{eg}(E_a)$
around the average trap field $E_a$ in the presence of the microwave dressing.
(c) Solid line: phase accumulated
by an $e/g$ coherence over a 40~ms period in presence of the microwave-dressing
showing a step structure (average step
size $\Delta \phi$) in phase with the atomic motion. Dashed line: a linear fit
(slope 6.9~rad/s) over the total time period of 1 s highlighting the
approximately linear evolution.}
\end{figure}

Similar broadenings are observed in optical dipole traps for
ground state atoms, where the potential energy is also, in general,
level-dependent. In this case, well-chosen trapping laser wavelengths
minimize the effect \cite{TR_KATORI03}. In the circular atom trap,
we propose to use instead a microwave state-dressing.
The atom is illuminated with a microwave field at angular
frequency $\omega_0$ polarized parallel to $Oz$, detuned by
$\delta_0=\omega_{eg}(E_a)- \omega_0>0$ to the red of the $e/g$
transition Stark shifted by the average field $E_a$ [see figure
2(a)]. This microwave field is therefore $\pi$-polarized and compatible with the
boundary conditions set by the trap electrodes. It is assumed to be a
propagating guided wave. Its amplitude, independent of $z$ and
$x$, is maximal at the origin (corresponding Rabi frequency
$\Omega_0$) and varies sinusoidally with $y$, having nodes at
$y=\pm 1$~ cm.
To first approximation, this field couples $g$ to the Rydberg
state $i$ ($n=51$, $m=m_g=49$), which experiences a large linear
Stark effect [polarizability 1~MHz/(V/m)]. The Stark polarizability
of the `dressed' $g$ level is thus significantly modified. The
dressing field has a smaller effect on $e$ since the
$\pi$-polarized transition from $e$ up to the $n=52$ manifold is
far off-resonance. The Rabi frequency $\Omega_0$ and the detuning
$\delta_0$ of the dressing microwave can therefore be tailored to
cancel the first and to minimise the second order terms in the
expansion of $\omega_{eg}(E)$ around $E_a$. Note that the dressed
$g$ level can still only decay through the emission of a
$\sigma$-polarized microwave photon. The exact shift of levels $g$
and $e$ for an atom located at $\bm{r}$ is calculated by an
explicit diagonalization of the atom-field hamiltonian, involving
8 hydrogenic manifolds, taking into account the local dressing
amplitude, time-dependent electric field and angle $\theta$
between $Oz$ and $\bm{E}(\bm{r},t)$. 
Note that a non-zero $\theta$
value causes the atom to experience a reduced component of the
dressing microwave parallel to $\bm{E}(\bm{r},t)$, and hence $\bm{u}$, and,
accordingly, a small microwave field orthogonal to $\bm{u}$. We have
checked that the latter
plays no important role, being far off-resonance from any allowed
transition originating from $g$ or $e$.

The optimal parameters
$\Omega_0$ and $\delta_0$ have reasonable values: $2\pi \times
200.000$ MHz and $2\pi \times 556.230$ MHz respectively. The
`dressed' transition frequency, shown in fig.~2(b), varies by only
$\sim$10~Hz over the $\pm$1~V/m electric field range explored by
an atom for $T_0=0.3\ \mu$K. The dressing reduces the transition
broadening by over 3 orders of magnitude. 
A quantitative confirmation of this insight must take into account the
modulation
of the atom-dressing field coupling due to the time-varying angle
$\theta$ contributing to a residual broadening of the line
(of about 0.5~Hz in our case).
Note that the adverse influence of stray electric fields (for instance those due
to patch effect) will also be dramatically reduced.
Based on the results of Ref.~\cite{VAHIDPATCH}, we estimate a patch-induced
field to have an average amplitude of 8~mV/m with a dispersion of 0.4~mV/m over
the trap volume. 

In order to estimate the coherence decay time $T_2$ we simulate a
Ramsey interferometry experiment. The atom undergoes two short
microwave $\pi/2$ pulses, resonant on the $e/g$ transition and
separated by a long delay $t$. We compute many trajectories
originating from an atomic cloud of size 0.3 $\mu$m, centred at
$O$, at temperature $T_0=0.3\ \mu$K and receiving a single optical
recoil along $Ox$ during Rydberg excitation. The phase $\phi(t)$
accumulated by an $e/g$ coherence is integrated over each
trajectory. Fig.~2(c) represents this phase evolution for a given
trajectory over a 40~ms time interval.
Most of the dephasing occurs at the turning points of the atomic
macromotion, where $\theta$ and $E-E_a$ are maximum. At these points the
influence of the micromotion, being at a higher frequency and of smaller
amplitude, averages, resulting in an accumulated dephasing $\Delta \phi$, almost
identical from one turning point to another. Therefore, over times long compared
to the trap period, the resulting dephasing evolves linearly.
%
%
The final fringe contrast is $C(t)=\left[  \overline{\cos \phi(t)}^2
+\overline{\sin\phi(t)}^2\right] ^{1/2}$ (where the bar denotes an average over
all trajectories). The time evolution of $C$, shown in figure 3,
provides an estimate of $T_2$ around 25 ms.

As noted earlier, the phase drift is almost perfectly linear with
time for all trapped trajectories. The phase spreading can
therefore be combated using an echo technique reminiscent of
photon echoes and of NMR refocusing schemes. Coherence preserving
echoes have also been tested for trapped ground-state atoms or
ions \cite{TR_ECHOATOMS}. At a time $t_{\pi}=0.5$~s after the first
Ramsey pulse, we `apply' a $\pi$-pulse on the $e/g$ transition,
exchanging the populations of the two states and hence changing
the sign of the accumulated phase (we mimic pulse imperfections by
a 10\% gaussian dispersion of the rotation angle). During the
subsequent evolution, the phase drift continues as before. The
phase of each trajectory thus returns towards zero. At time
$t=2t_{\pi}=1$ s, all phases are zero within an uncertainty of the
order of the average phase step amplitude $\Delta\phi$. Figure 3
also presents the contrast $C$ obtained under these conditions as
a function of $t$. It increases sharply up to 82\% around
$t=2t_{\pi}=1$ s. This very high contrast corresponds to an
effective $T_2=5.0$ s. Even for atoms at $T_0=1\ \mu$K, we obtain
$C(2t_{\pi})=$57.3\%. More complex echo sequences can be envisaged
to improve the final Ramsey fringe contrast. Ideal $\pi$ pulses
repeated at shorter time intervals can maintain the coherence over
time scales in the minute range.

\begin{figure}
\includegraphics[width=7cm]{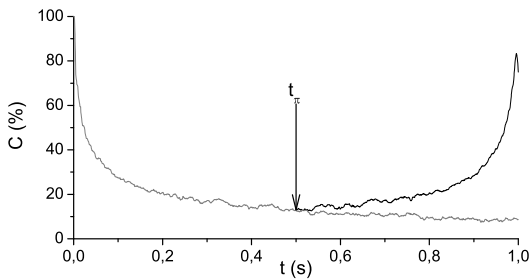}
\caption{\label{Fig:echo} Simulated Ramsey fringe contrast, $C$ as a
function of the time interval $t$ between two $\pi/2$ pulses. Grey line:
contrast decay without echo, averaged over 10000 trajectories. The contrast
undergoes a non-exponential decay, falling to 50\% in 24 ms, and to 13\% in 0.5
s.
Black line: $C$ versus $t$ for the same trajectories with a $\pi$-pulse at
$t_{\pi}$=0.5 s. The sharp contrast revival at $t$=1 s peaks at 82\%.}
\end{figure}

We have shown that the combination of S.E. inhibition, Stark
trapping, microwave dressing and echo techniques makes it possible
to manipulate Rydberg atoms coherently over long time intervals.
The Stark trap presents advantages as compared to magnetic traps,
which do not possess an equivalent coherence preservation scheme
to combat the dephasing due to the differential Zeeman effect of e
and g. Traps based on ponderomotive
forces \cite{TR_PONDEROMOTIVE00}, requiring large laser powers, are
hardly compatible with a cryogenic environment. 

We have also designed electric traps with smaller electrodes and checked that
they are compatible with the coherence preservation scheme. These would bring
the trapped atom much closer to the electrode surface which can be of interest
for atom-surface studies. Elaborating on our scheme, one can design a variety of
different structures such as waveguides or `conveyor belts'. These would
operate via the use of electrode arrays, easily produced by
lithographic techniques, and a proper commutation of the voltages
applied to them. Field-ionization Rydberg atom detectors could
also be realized `on chip', using the accelerated electron to
trigger a superconducting to normal transition in a mesoscopic
wire \cite{TR_DAY03}. These `everlasting' giant Rydberg atoms open a
wealth of fascinating perspectives for fundamental studies. High
spectroscopic resolution, at the Hertz level, makes it possible to
study atom/surface interactions of the Casimir type, at
millimetre-range distances \cite{ENS_HOUCHES90,QC_VDWCASIMIR}.
Atom/atom \cite{QI_LUKINDIPOLEGATE00} interaction at long range,
possibly mediated by a superconducting
line \cite{QI_LUKINTRANSMISSIONLINE04}, could also be investigated.
The coherent interaction of atoms with superconducting quantum
circuits on the chip surface is also a promising avenue of
inquiry. Cavity QED experiments could be realized with high-Q
planar transmission-line
cavities \cite{TR_SCHOELKOPFTRANSMISSIONLINE04}. Atom/atom
entanglement could be generated through cavity-assisted
collisions \cite{ENS_COLLISION}, the scheme being, in principle,
extendable to much more complex architectures for quantum
information processing.

\begin{acknowledgments}
Laboratoire Kastler Brossel is a laboratory of Universit\'{e}
Pierre et Marie Curie and ENS, associated to CNRS (UMR 8552). We
acknowledge support of the European Community (QUEST and QGates
projects), of the Japan Science and Technology corporation
(International Cooperative Research Project~: Quantum
Entanglement).
\end{acknowledgments}



\end{document}